\documentclass[11pt,a4paper]{article}
\pdfoutput=1
\usepackage{jinstpub}
\usepackage{subcaption}
\usepackage[sort&compress]{natbib}

\begin{document}

\title{The CENNS-10 Liquid Argon Detector \\ to measure CEvNS at the Spallation Neutron Source}
\proceeding{LIDINE 2017: Light Detection in Noble Elements\\
22-24 September 2017\\
SLAC National Accelerator Laboratory}

\author[a]{R. Tayloe, }
\affiliation[a]{Department of Physics, Indiana University, \\ 
Bloomington, IN 47405, USA}
\emailAdd{rtayloe@indiana.edu}

\author[b]{for the COHERENT collaboration}
\affiliation[b]{https://sites.duke.edu/coherent/collaboration/}

\abstract{
The COHERENT collaboration is deploying a suite of low-energy detectors in a low-background corridor of the
ORNL Spallation Neutron Source (SNS) to measure coherent elastic neutrino-nucleus scattering (CEvNS) on an array
of nuclear targets employing different detector technologies.   A measurement of CEvNS on different nuclei will test the  $N^2$-dependence of the CEvNS cross section and further the physics reach of the COHERENT effort. The first step of this program has been realized recently with the observation of CEvNS in a 14.6 kg CsI detector. Operation and deployment of Ge and NaI detectors are also underway.   A 22 kg, single-phase, liquid argon detector (CENNS-10) started  data-taking in Dec. 2016 and will provide results on CEvNS from a lighter nucleus.  Initial results indicate that light output, pulse-shape discrimination, and background suppression are sufficient for a measurement of CEvNS on argon. 
}

\maketitle

\section{COHERENT experiment}
\begin{figure}
\centering
\hfill
\begin{subfigure}[b]{0.34\textwidth}
\includegraphics[width=\textwidth]{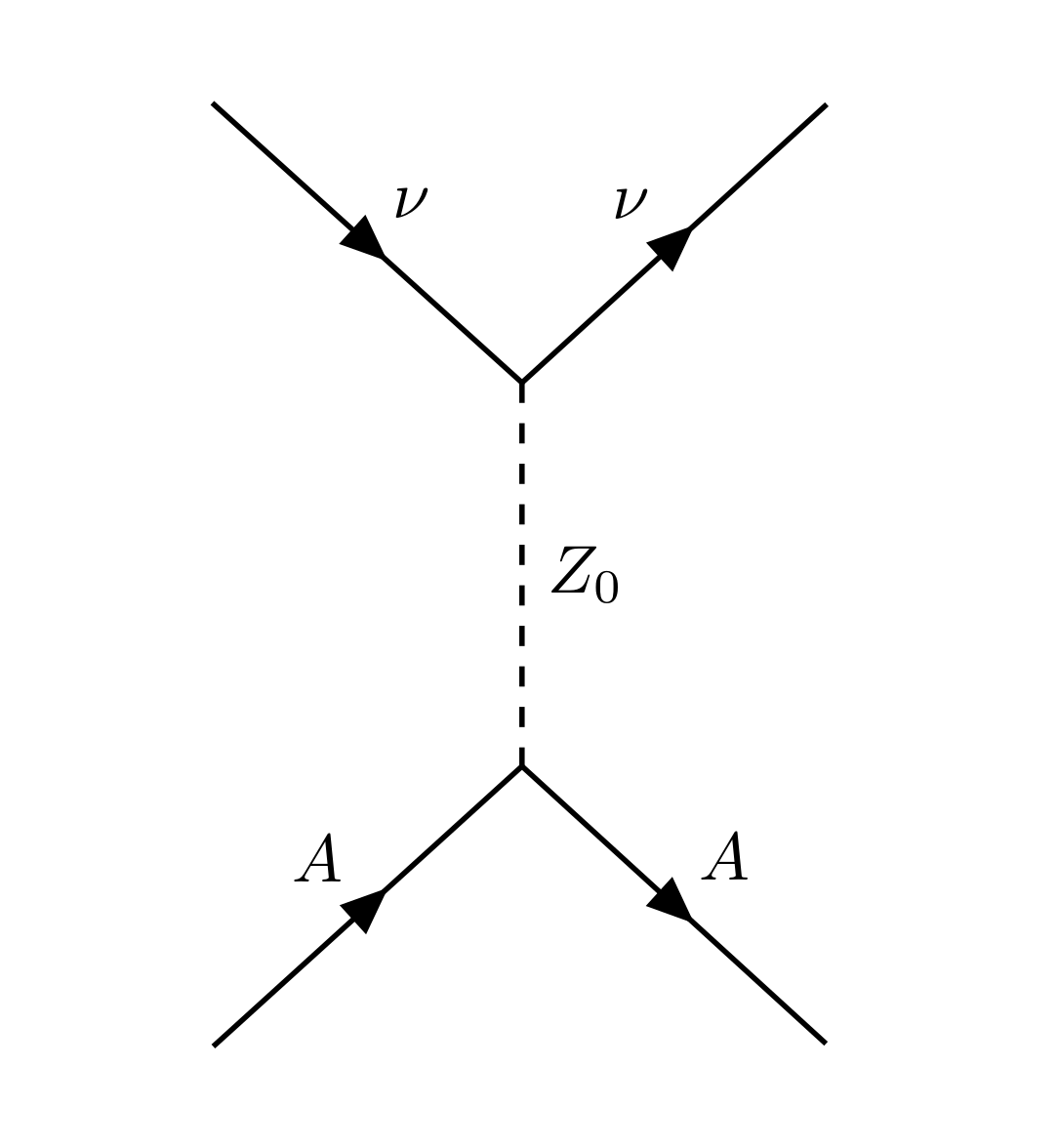}
\caption{}\label{fig:nuAdiagram}
\end{subfigure}
\hfill
\begin{subfigure}[b]{0.6\textwidth}
\includegraphics[width=\textwidth]{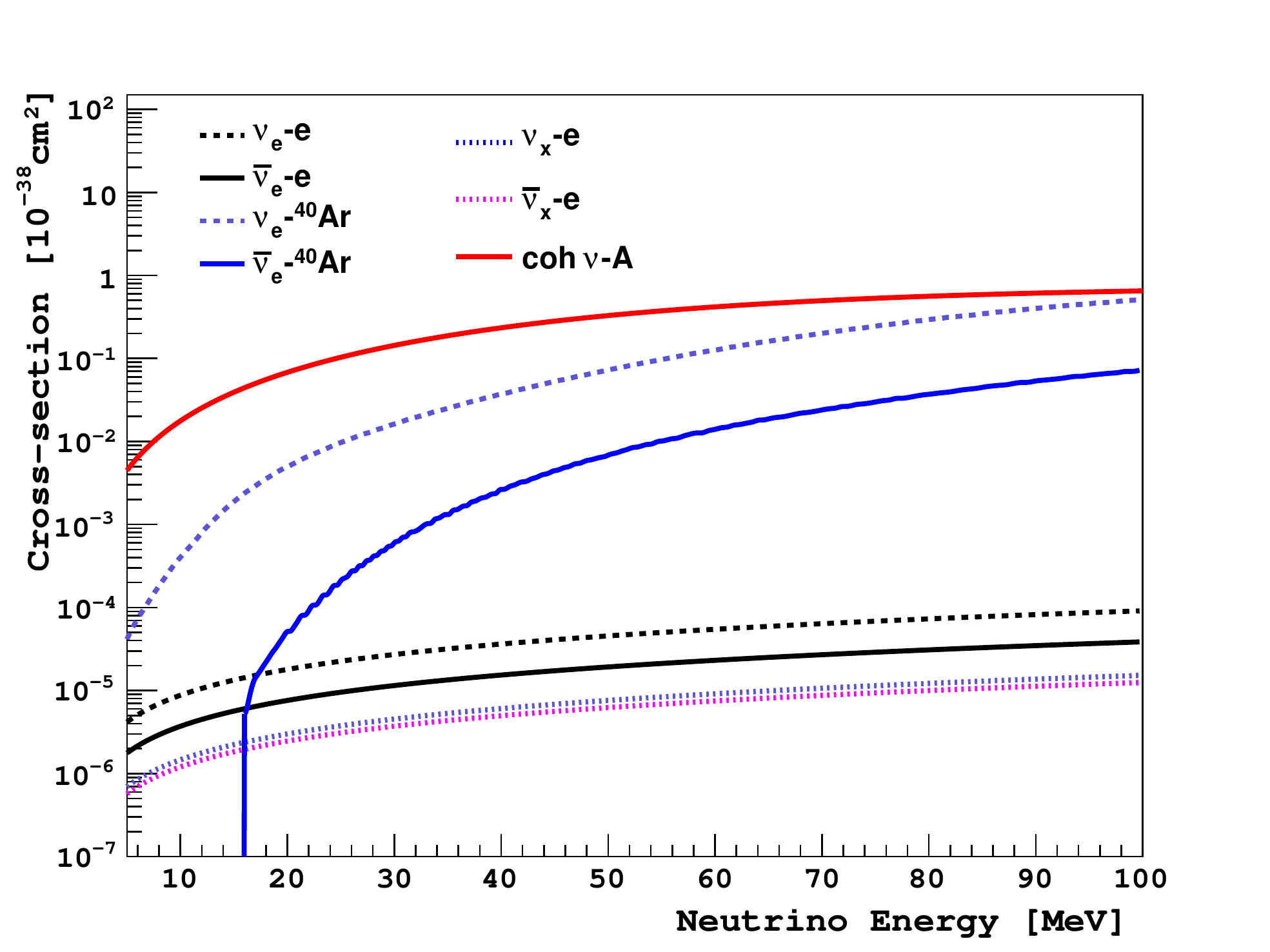}
\caption{}\label{fig:nuAxsArgon}
\end{subfigure}
\caption{$\nu A$ scattering:  (a) Feynman diagram~\cite{Ellis:2016jkw} for the process and (b) $\nu A$ and $\nu e$ scattering cross sections as a function of neutrino energy. The CEvNS process is shown as a red line and is compared to competing $\nu$ processes in an argon target.}\label{fig:nuA}
\hfill
\end{figure}

Coherent elastic neutrino-nucleus scattering (CEvNS) is the interaction of a neutrino ($\nu$) with a nucleus (A) (Fig.~\ref{fig:nuAdiagram}) with sufficiently low momentum transfer that the nucleus remains in the original ground state.  In that case the neutrino sees the nucleus as a whole and the collection of individual nucleons behaves ``coherently'' resulting in a large (for neutrinos) interaction probability that is the largest of competing neutrino processes in this energy region (Fig.~\ref{fig:nuAxsArgon}). In addition, the weak-interaction charge of the proton is small compared to that of the neutron so the CEvNS rate is predicted to be insensitive to protons and depends upon $N^2$, the square of the number of neutrons. Coherence dictates that the associated wavelength of the momentum transfer is large compared to the size of the nucleus.  For medium-sized nuclei (like argon), this constrains the initial neutrino energy to $\cal{O}$(50~MeV) and yields nuclear recoil energies of $\cal{O}$(10~keV).  This is challengingly low for typical detector technologies and the main reason that the CEvNS has remained undetected until only recently~\cite{Akimov:2017ade}, $\approx 50$ years after its prediction~\cite{Freedman:1973yd}. 

The newly-discovered CEvNS process provides a high rate of events in a modest-sized detector,  
is sensitive to the details of the distribution of neutrons in a nucleus, and has a well-understood standard model prediction.  This system is ideal to provide new physical insight. For example, is the interaction rate and recoil distribution as predicted by the standard model?  The answer will have large implications in understanding supernovae~\cite{Wilson:1974zz,Horowitz:2004pv}, non-standard interactions~\cite{Barranco:2007tz,Coloma:2017ncl,Liao:2017uzy}, and the nuclear physics of neutron distributions~\cite{Patton:2012jr,Cadeddu:2017etk}.  

The COHERENT experiment~\cite{Akimov:2015nza} was designed to detect CEvNS events and to demonstrate the predicted $N^2$-dependence of the process by employing a variety of detectors with different nuclear targets.  In addition, the different detector technologies have different systematic errors, so a CEvNS signal in all will offer a more precise and robust measurement.

Additionally, these neutrino measurements are also important in the current search for dark matter (DM) in two ways.  \textbf{1)} So-called ``direct-searches'' for DM are sensitive to nuclear recoils that may occur due to DM interactions in the detector.   These recoils look just like CEvNS interactions from solar or atmospheric neutrinos which will soon be the dominant background for the latest DM direct-searches (e.g.~Refs~\cite{Cui:2017nnn,Aprile:2017iyp}).  Our proposed measurement will independently measure this DM-search background rate using the same detector technology.  \textbf{2)} An $\cal{O}$(1-ton) LAr detector will be sensitive to DM produced in the neutrino beam as predicted in ``vector-portal'' models~\cite{deNiverville:2015mwa} of DM.  These alternative theories of DM are becoming more compelling as direct-searches for DM yield null results.

\begin{figure}
\centering
\hfill
\begin{subfigure}[b]{0.58\textwidth}
\includegraphics[width=\textwidth]{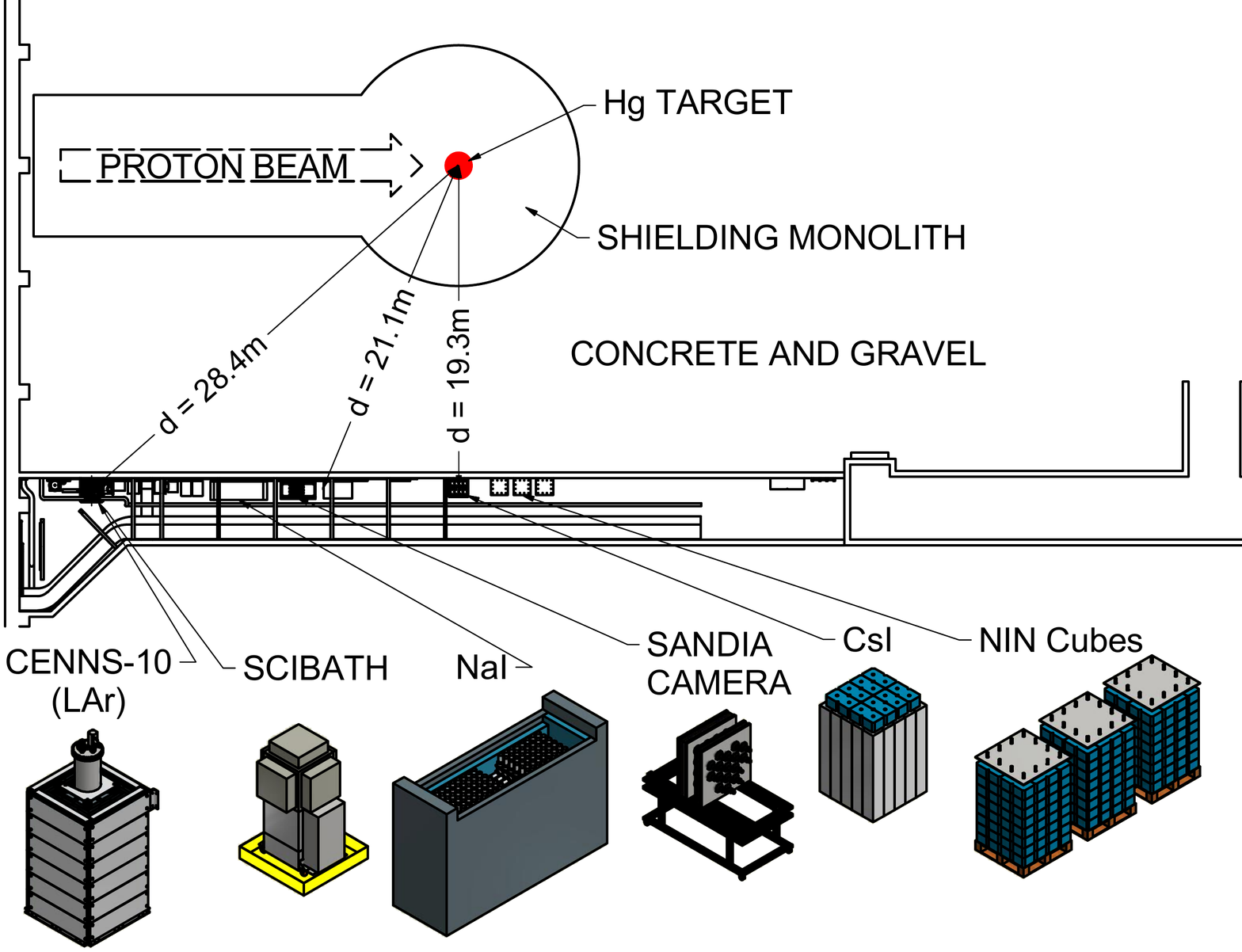}
\caption{} \label{fig:nuAlley}
\end{subfigure}
\hfill
\begin{subfigure}[b]{0.38\textwidth}
\includegraphics[width=\textwidth]{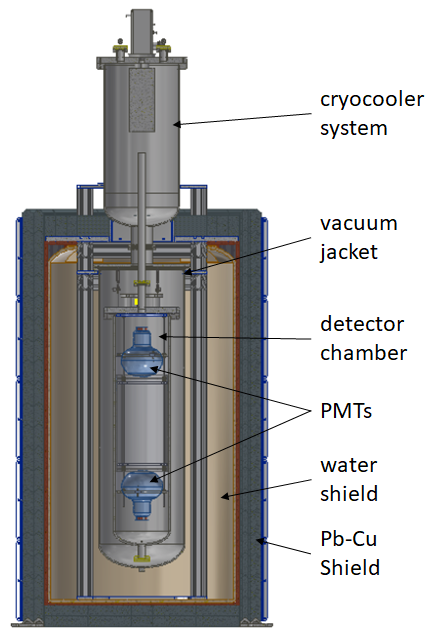}
\caption{} \label{fig:CENNS10detector}
\end{subfigure}
\hfill
\caption{The COHERENT experiment at the SNS: (a) $\nu$-alley layout and (b) a section view of the CENNS-10 detector.}
\label{fig:COHERENT}
\end{figure}

The COHERENT experiment~\cite{Akimov:2015nza} is using a suite of detectors at the Oak Ridge National Laboratory (ORNL) Spallation Neutron Source (SNS) in a phased approach to first discover, then measure the CEvNS process on a variety of detectors employing different nuclear targets.  The detectors are deployed in a basement hallway, ``$\nu$-alley'' (Fig.~\ref{fig:nuAlley}), that has been determined to be, from a thorough measurement campaign utilizing $\gamma$ and neutron detectors, sufficiently low in background for the CEvNS measurements.  The detectors are arrayed along the corridor at distances 20-29~m from the intense 1~MW SNS neutron production target that provides a high flux of $\pi$-decay neutrinos up to $\approx 50$~MeV, ideal for CEvNS interaction. Additionally, the beam is pulsed, providing an $\approx 1\times10^{-5}$ steady-state background rejection factor.  As mentioned above, the first milestone in this effort was reached recently with the $6.7\sigma$ discovery, of the CEvNS process in $\approx$ 1.5 years of running a Cesium Iodide (CsI) detector. 

\section{CENNS-10 LAr detector}
The CENNS-10 single-phase liquid argon (LAr) detector was designed and built at Fermilab for the CENNS effort at Fermilab~\cite{Brice:2013fwa}.  After initially commissioning at Fermilab, it was transported to Indiana University for testing with neutrons.  An opportunity arose to measure CEvNS with this detector on an accelerated timescale, and the detector was reconfigured and shipped to ORNL for installation at the SNS in late 2016.  It ran with beam from December 2016 to May 2017 (phase-1) and was then upgraded in June 2017, collecting beam data from July-December 2017 (phase-2).  

The CENNS-10 detector (Fig.~\ref{fig:CENNS10detector}) consists of a cylindrical 56.7~L LAr chamber, viewed by two opposing 8'' Hamamatsu R5912-02MOD photomultiplier tubes (PMT) that are read out with a (12-bit, 250~MHz) CAEN 1720 digitizer.   In phase-1 running the sides were lined with an acrylic cylinder,  spray-coated with tetraphenyl butadiene (TPB) in dichloromethane to produce a 0.2~mg/cm$^2$ TPB layer, 
to wavelength-shift the 128~nm LAr scintillation light into the visible range.  The cylinder was 27~cm diameter by 46~cm in length and backed by a Teflon sheet.   Acrylic disks with an evaporatively-coated TPB layer of the same thickness were mounted at the ends of the cylinder in front of the PMTs forming a fiducial volume of 32~kg. 

The light production in this phase-1 arrangement was inadequate, so for phase-2 running the acrylic cylinder was replaced with a 21~cm diameter TPB-coated Teflon cylinder and the acrylic end disks were removed and replaced with 2 new PMTs with directly-coated TPB on the front surface.  The Teflon cylinder was evaporatively coated at ORNL and the PMT by Intlvac Thin Film, both to 0.2~mg/cm$^2$.
This new configuration formed a smaller active volume of 22~kg, but the light output was increased by about a factor of 5 (as shown below).

The LAr is liquefied and maintained at temperature with a Cryomech PT-90  pulse-tube cryorefrigerator.  The cold head and associated plumbing is mounted above the detector chamber and both the detector and cryocooler are enclosed in an insulating stainless steel vacuum vessel maintained at $1\times10^{-6}$~Torr.  Boil-off argon gas from the detector is routed through a heat exchanger in the cryocooler and then to the cold head for re-liquification.  The Ar gas system in the cryocooler is connected to an external gas handling system where it is circulated at $\approx 5$~slpm through a SAES Zr getter operating at $400^\circ$~C to remove residual gases providing $\approx 1$~ppm purity.  There is also a path for Ar gas back to the cold head where it is cooled and condensed into the detector during steady-state running or for the initial fill with Argon gas. 

The insulation vessel is immersed in a plastic tank providing an $\approx 40$~cm layer of water surrounded by 0.6~cm copper, 10~cm lead, and stabilized with exterior 0.6~cm aluminum plates.  This shielding combination is designed to reduce external, steady-state $\gamma$ backgrounds along with the low flux of beam-related neutrons.

\begin{figure}
\centering
\hfill
\begin{subfigure}[b]{0.48\textwidth}
\includegraphics[width=\textwidth]{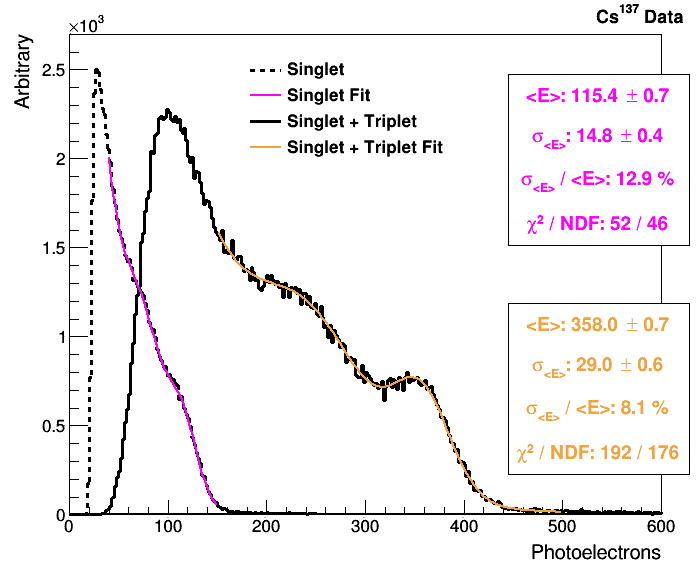}
\caption{} \label{fig:CENNS10ph1Cs137spectrum}
\end{subfigure}
\hfill
\begin{subfigure}[b]{0.48\textwidth}
\includegraphics[width=\textwidth]{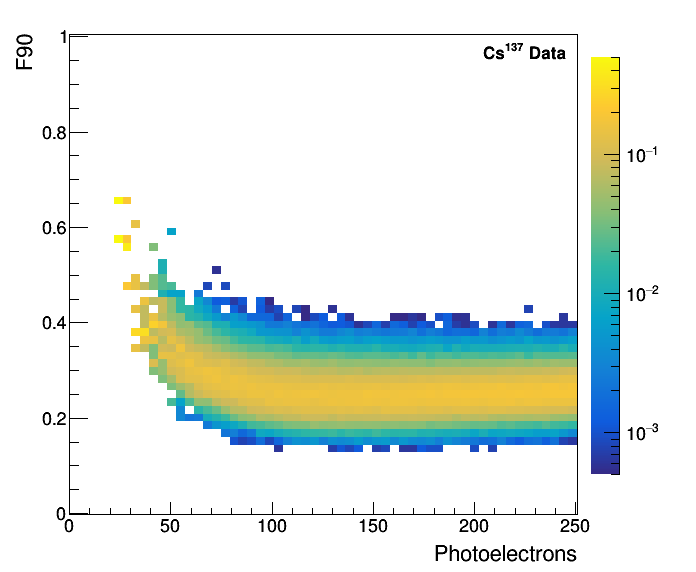}
\caption{} \label{fig:CENNS10ph1Cs137PSD}
\end{subfigure}
\hfill
\\
\hfill
\begin{subfigure}[b]{0.48\textwidth}
\includegraphics[width=\textwidth]{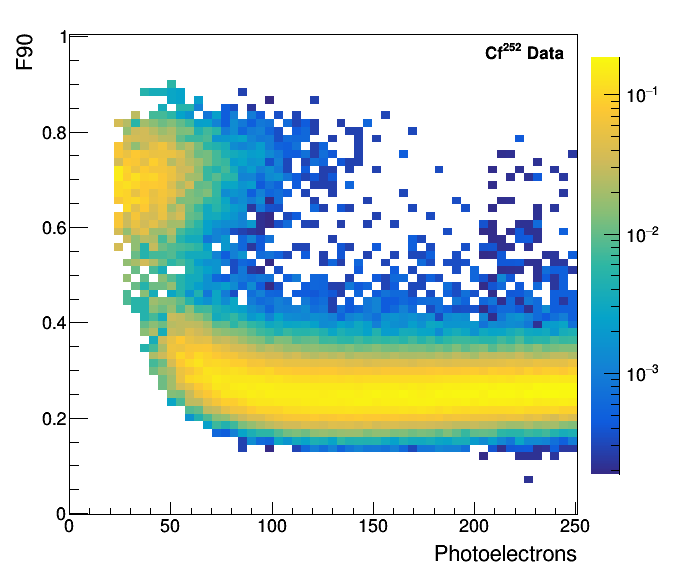}
\caption{} \label{fig:CENNS10ph1Cf252PSD}
\end{subfigure}
\hfill
\begin{subfigure}[b]{0.48\textwidth}
\includegraphics[width=\textwidth]{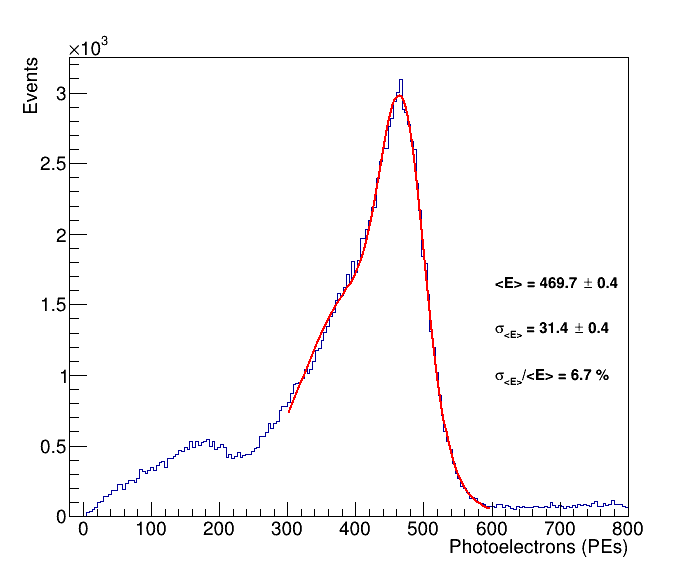}
\caption{} \label{fig:CENNS10ph1Co57spectrum}
\end{subfigure}
\hfill
\caption{Initial data distributions from phase-1 (a-c) and phase-2 (d) data:  (a) Photoelectron (PE) spectrum for $^{137}$Cs;  (b) F90 pulse-shape variable for $^{137}$Cs; (c) F90 for $^{252}$Cf; and phase-2 data, $^{57}$Co PE spectrum. Note that these are preliminary results.  In (a), the singlet (prompt) and total PE spectra are displayed and fit to a $^{137}$Cs spectrum separately.  Fit parameters are the mean, absolute/relative  widths of the fit to the 662~keV photopeak and the $\chi^2/NDF$ for the fit.}
\end{figure}

\section{Initial results and outlook}
For phase-1 running of the CENNS-10 detector, initial results indicated that the wavelength-shifting and light collection efficiency would not provide a sufficiently low energy threshold for a measurement of the CEvNS process.  This is most likely due to low efficiency for TPB reemission from the TPB-coated acrylic on the side walls of the detector.  However, important data on the detector operation and energy response was collected.   Single photoelectron (PE) response of the phototube/readout electronics was determined via the injection of small light pulses from an external blue LED coupled with a fiber optic cable.  The number of PE for a given energy could then be determined from the detector response to $\gamma$ sources placed just outside the vacuum insulation volume.  The primary source used for the phase-1 configuration was $^{137}$Cs (662~keV). The PMT waveforms for calibration events were fit to provide the prompt (``singlet'') and delayed (``triplet'') argon scintillation components.  The resulting calibration spectra are shown in Fig.~\ref{fig:CENNS10ph1Cs137spectrum} and indicate $\approx 0.5$ detected PE/keVee for this phase-1 configuration. 

Standard, atmospheric-derived argon contains $\approx 1$ Bq/kg of $^{39}$Ar that $\beta$-decays with a 565~keV endpoint.  It is a constant-rate background that may be subtracted from any beam-related signal, however, if the rate is too large, statistical fluctuations will be overwhelming.  These $\beta$-decay events can be discriminated from CEvNS-event nuclear recoils via the differences in triplet-singlet light amplitude between the two event types~\cite{Boulay:2006mb}.  This is accomplished by examining the fraction of light in the first 90~nsec of an event pulse (``F90'').  This quantity will be larger for CEvNS-event nuclear recoils as the triplet light is suppressed, compared to that for $e/\gamma$ events.   Figures~\ref{fig:CENNS10ph1Cs137PSD},\ref{fig:CENNS10ph1Cf252PSD} show the discrimination, for the phase-1 configuration, between $^{137}$Cs $\gamma$ events with an F90 value of $\approx 0.3$ and neutrons from a $^{252}$Cf source with F90 $\approx 0.7$.

The phase-2 upgrade of CENNS-10 in Summer '17 was expected to increase the light output significantly because of the improved wavelength-shifting and reflection of the Teflon sides compared to that of acrylic.  Initial examination of the data shows that to be realized.  The single-PE calibration is performed weekly and is reasonably stable over few-months timescales.  Figure~\ref{fig:CENNS10ph1Co57spectrum} shows the PE distribution of total (singlet + triplet) light in $^{57}$Co (122~keV $\gamma$) events.  This indicates an improvement in light yield of 4-5$\times$ from that in phase-1.   Current analysis also shows that the PSD is adequate for this data as well.  This will allow sensitivity to nuclear recoil events down to $\approx 20$~keV, sufficient for collection of $\approx 150$ CEvNS events in 1 year of running at the SNS and a first measurement of CEvNS with an argon target.

\bibliographystyle{JHEP}
\bibliography{RTLIDINE17}

\end{document}